\documentclass[runningheads]{llncs}
\usepackage[T1]{fontenc}
\usepackage{graphicx}
\usepackage{soul}
\usepackage{url}
\usepackage[utf8]{inputenc}
\usepackage{amsmath}
\usepackage{booktabs}
\usepackage{algorithm}
\usepackage{algorithmic}
\usepackage[switch]{lineno}

\usepackage{wrapfig}
\usepackage{float}

\usepackage{amsmath}
\usepackage{makecell}
\usepackage{multirow}
\usepackage{arydshln}
\usepackage{booktabs}
\usepackage{xcolor}
\usepackage{xspace}
\usepackage{listings}
\usepackage{array}

\usepackage{multirow}
\usepackage{graphicx}
\usepackage{makecell}
\usepackage{tabularx} 
\usepackage{marvosym}
\usepackage{wasysym}

\newcolumntype{C}{>{\centering\arraybackslash}X}

\lstdefinelanguage{TXT}{
  basicstyle=\ttfamily\fontsize{9}{10}\selectfont,
  sensitive=true,
  morekeywords={true,false,null},
  keywordstyle=\color{blue},
  commentstyle=\color{green},
  showstringspaces=false,
  breaklines=true,
  columns=flexible,
  frame=single,
  backgroundcolor=\color{gray!8},
  breakindent=0pt,         
  breakautoindent=false,   
}

\lstdefinelanguage{JSON}{
  basicstyle=\ttfamily\fontsize{9}{10}\selectfont,
  sensitive=true,       
  morekeywords={true,false,null}, 
  keywordstyle=\color{blue}, 
  commentstyle=\color{green}, 
  showstringspaces=false, 
  breaklines=true, 
  columns=flexible, 
  frame=single, 
  backgroundcolor=\color{gray!8}, 
}

\newcommand{\sys}{{ORANGE}\xspace}

\newcommand{\sql}[1]{\texttt{\fontfamily{pcr}\selectfont {#1}}}

\begin{document}
%

\title{ORANGE: An Online Reflection ANd GEneration framework with Domain Knowledge for Text-to-SQL}

\author{Yiwen Jiao\inst{1}\textsuperscript{*} \and
Tonghui Ren\inst{2}\textsuperscript{*} \and
Yuche Gao\inst{3}\textsuperscript{*} \and
Zhenying He\inst{1} \textsuperscript{\Letter} \and
Yinan Jing\inst{1} \and
Kai Zhang\inst{1} \and
X.Sean Wang\inst{1}}

\authorrunning{Jiao et al.}

\institute{Fudan University, China \\
    \email{ywjiao24@m.fudan.edu.cn, \{zhenying,jingyn,zhangk,xywangCS\}@fudan.edu.cn} \and
    Tencent Cloud, China
    \email{rayeeren@tencent.com} \and
    University of Cambridge, United Kingdom
    \email{yg473@cam.ac.uk}}
    
\maketitle

\makeatletter
\renewcommand\@makefntext[1]{#1}
\makeatother

\footnotetext{* Equal contribution \hspace{1em} \textsuperscript{\Letter} Corresponding author}

%

\begin{abstract}
Large Language Models (LLMs) have demonstrated remarkable progress in translating natural language to SQL, but a significant semantic gap persists between their general knowledge and domain-specific semantics of databases.
Historical translation logs constitute a rich source of this missing in-domain knowledge, where SQL queries inherently encapsulate real-world usage patterns of database schema. Existing methods primarily enhance the reasoning process for individual translations but fail to accumulate in-domain knowledge from past translations. We introduce \sys, an online self-evolutionary framework that constructs database-specific knowledge bases by parsing SQL queries from translation logs. By accumulating in-domain knowledge that contains schema and data semantics, \sys progressively reduces the semantic gap and enhances the accuracy of subsequent SQL translations. To ensure reliability, we propose a novel nested Chain-of-Thought SQL-to-Text strategy with tuple-semantic tracking, which reduces semantic errors during knowledge generation. Experiments on multiple benchmarks confirm the practicality of \sys, demonstrating its effectiveness for real-world Text-to-SQL deployment, particularly in handling complex and domain-specific queries.



\end{abstract}

\section{Introduction}


Large Language Models (LLMs) have demonstrated remarkable capabilities in translating natural language questions into executable SQL queries, significantly lowering the barrier for non-technical users to interact with databases~\cite{CHASE,DIN,CHESS,MAC}. However, a critical challenge is the semantic gap between the general-purpose knowledge embedded in LLMs and the domain-specific semantics of the target database schema.

\begin{figure}[tb]
    \vspace{-5pt}
    \centering
    \includegraphics[width=0.95\linewidth]{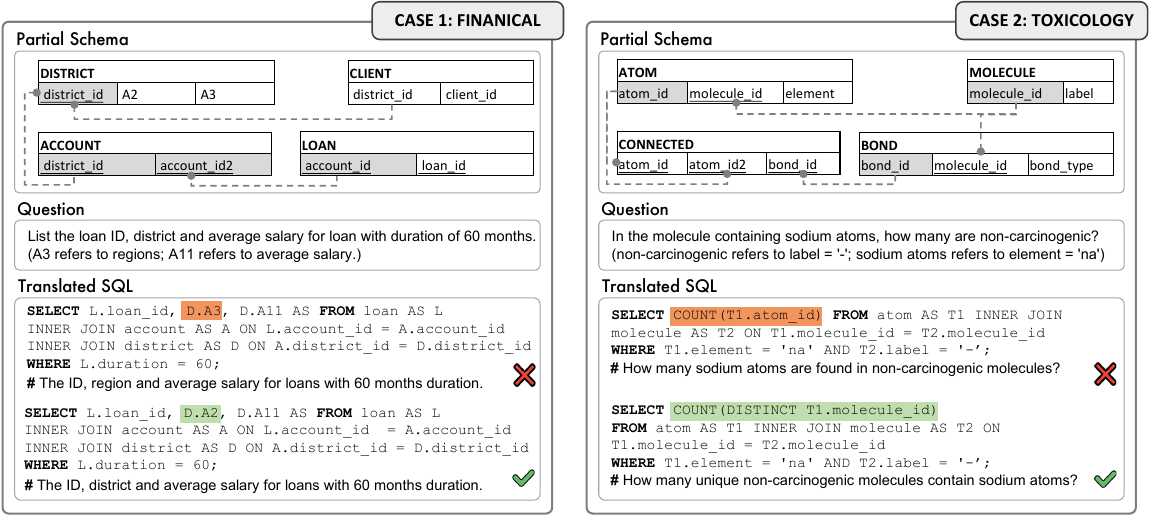}
    \caption{A Text-to-SQL example.}
    \label{fig_example}
    \vspace*{-15pt} 
\end{figure}

Unlike general text generation, Text-to-SQL requires in-domain reasoning that often goes beyond the database schema information alone. Crucial knowledge about schema semantics, value distributions, and business logic is difficult to infer from limited schema definitions and is frequently absent from user questions. Consequently, this semantic gap leads to systematic errors, such as misinterpretation of column meanings or operational intent, resulting in SQL queries that are syntactically valid but semantically incorrect~\cite{chen2025reliable,knowledge-to-sql}. As shown in Figure~\ref{fig_example}, Case 1 illustrates that when processing the query \textit{average salary in each district}, a model lacking in-domain knowledge erroneously uses the \textit{regions} column \sql{A3} instead of the correct \textit{districts} column \sql{A2}. In Case 2, the \sql{ATOM INNER JOIN MOLECULE} operation produces an intermediate table, where each tuple no longer represents the intended molecule, but instead an atom associated with its molecule. This shift in tuple semantics  causes aggregation errors, such as incorrectly using \sql{COUNT(ATOM.id)} to count atoms rather than the correct operation \sql{COUNT(DISTINCT ATOM.molecule\_id)} to count molecules.

To effectively utilize LLMs for Text-to-SQL, the dominant paradigm focuses on enhancing the reasoning process of LLMs through techniques such as optimized prompting~\cite{PET-SQL,DIN,DEA-SQL}, task decomposition~\cite{zhou2022least}, refinement~\cite{SQLFixAgent,SelfDebug}, self-consistency voting~\cite{CHESS,MAC}, and test-time scaling~\cite{xiyan,CHASE}. While these methods improve the SQL translation performance, they lack a mechanism to acquire and accumulate domain-specific insights from the target database. Each query is treated as an isolated task, thus failing to leverage past translation experiences. Recent studies highlight the importance of in-domain knowledge and attempt to incorporate it by generating synthetic domain-specific question or leveraging historical logs~\cite{chen2025reliable,LPE-SQL,CHASE}. However, synthetic data often misalign with real user intents and can introduce hallucinations, while log-based approaches typically require extensive manual annotation.

A more desirable solution is to create a Text-to-SQL system capable of using its translation logs to achieve evolution without human intervention~\cite{wang2024improving,zhang2024sqlfuse}. In this paper, we present \sys (\textbf{O}nline \textbf{R}eflection \textbf{AN}d \textbf{GE}neration), a self-evolutionary framework that parses and stores validated knowledge into  database-specific memory, which is then leveraged for subsequent in-context learning process. \sys consists of three core components: \textbf{Knowledge Decomposition}, \textbf{Knowledge Validation}, and knowledge-enhanced \textbf{Text-to-SQL Translation}. SQL operations such as \sql{JOIN} and \sql{Aggregation} alter tuple semantics, resulting in complex reasoning~\cite{GAR}. To accurately capture these shifts, we propose a nested Chain-of-Thought (CoT) approach that decomposes SQL queries into subcomponents for progressive annotation and explicitly tracks tuple-semantic shifting. This design ensures that the generated knowledge units faithfully reflect the database-specific semantics. Unlike prior methods, \sys relies solely on SQL queries from translation logs without original user queries, annotating the semantics of knowledge through tuple-semantic tracking and constructing a database-specific knowledge base. These verified in-domain knowledge can be reused to guide future predictions and improve the translation accuracy of \sys over time without manual intervention.

We conduct extensive experiments on three benchmarks~\cite{SPIDER,BIRD,SCIENCE} to evaluate \sys. The results show consistent accuracy improvement over baselines, indicating the robustness of the proposed self-evolutionary framework for Text-to-SQL. Our contributions are as follows:

    
    

\vspace{-0.2cm}
\begin{enumerate}
    \item We propose a self-evolutionary Text-to-SQL paradigm that accumulates and reuses in-domain knowledge without human intervention.
    
    \item We introduce \sys, an online reflection and generation framework that constructs a reliable, domain-specific knowledge base from translation logs through nested CoT strategy with tuple-semantic tracking.
    
    \item We conduct extensive experiments to validate the effectiveness of \sys.
\end{enumerate}




    
    





\section{Methods}

\subsection{Methodology Overview}

Our work operates as a self-evolutionary framework that uses the translation logs to construct a reusable, in-domain knowledge base. To translate a natural language question $X$ into a target SQL query $Y$ based on the database schema $\mathcal{S}$, \sys parses the historical translated SQL queries \(\mathcal{C}\) and maintains a memory $\mathcal{M}$ of verified $k = (k_x, k_y)$, where each unit consists of a Text-SQL pair. The system evolves $\mathcal{M}$ through successive translations, progressively enriching its domain understanding.

\begin{figure*}[t]
\vspace{-5pt}
    \centering
    \includegraphics[width=\linewidth]{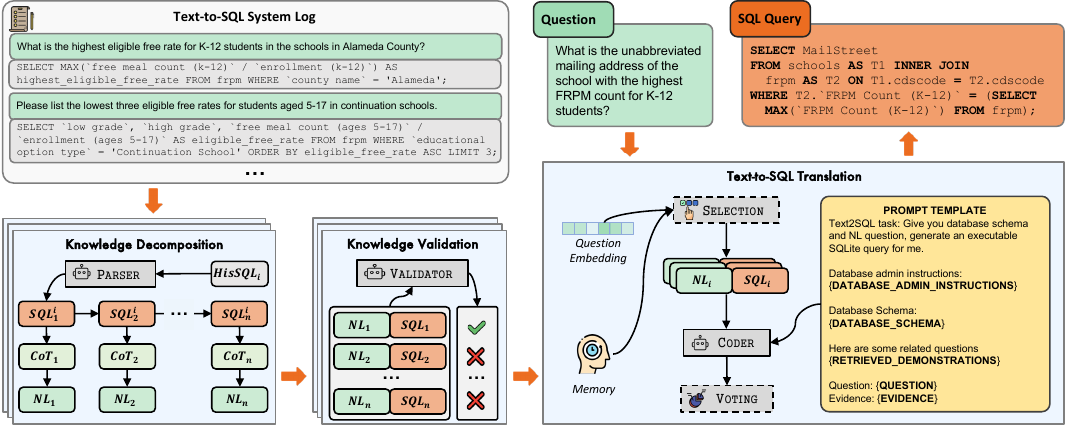}
    \caption{Overview of \sys.}
    \label{fig_overview}
    \vspace*{-15pt} 
\end{figure*}

As illustrated in Figure~\ref{fig_overview}, \sys operates with three stages: 
\vspace{-0.2cm}
\begin{enumerate}
    \item \textbf{Knowledge Decomposition}: Extracting generated SQL queries from the logs, the \textsc{Parser} adopts a nested CoT approach to parse SQL queries into knowledge units, each consisting of a semantically-aligned Text-SQL pair.
 
    \item \textbf{Knowledge Validation}: The \textsc{Validator} verifies the correctness of knowledge units through probability scores, only retaining the reliable ones in \(\mathcal{M}\).

    \item \textbf{Knowledge-Enhanced Translation}: The \textsc{Coder} performs SQL translation as an in-domain In-Context Learning (ICL) task. It retrieves relevant demonstrations from \(\mathcal{M}\), while incorporating multi-path generation with the self-consistency strategy to further enhance reliability.
    
\end{enumerate}





\vspace{-10pt}
\subsection{Knowledge Decomposition}\label{sec_knowledge_parsing}

This stage analyzes existing Text-to-SQL translation logs, which contain the generated candidate SQL queries for both the current question and previous tasks. For each question \(\mathcal{X}\), its candidate SQL queries \(\mathcal{C}\) are clustered by their execution results and are ranked by cluster size. We then select the first SQL query in each cluster to represent that cluster, denoted as \(\mathcal{C}'\).


For each SQL in \(\mathcal{C}'\), we employ a nested Chain-of-Thought (CoT) strategy that decomposes the SQL into sub-SQL components \(k_y\) and generates the corresponding question \(k_x\). The resulting knowledge units, each is defined as a Text-SQL pair \(k =(k_x, k_y)\), are then de-duplicated to serve as the output.


\vspace{-10pt}
\subsubsection{Nested CoT Reasoning}

We propose a nested Chain-of-Thought (CoT) strategy for SQL-to-SUBSQL-to-Text (SST) annotation. The approach uses an outer loop (SQL-to-SUBSQL) and an inner loop (SUBSQL-to-Text), which decomposes and interprets complex SQL step by step for better understanding.

In the outer loop, the \textsc{Parser} decomposes the SQL in a \textit{least-to-most} style. Simple queries with fewer operations form building blocks for more complex queries. This incremental approach allows partial reuse of earlier semantics. For each question \(X\), the outer loop decomposes the candidate SQL \(C_i\) as:

\vspace{-0.2cm}
\begin{equation}\small
P = \textsc{Parser}(C_i, S, X_e),
\end{equation}

where \(X_e\) is the evidence used in the original question. We only use \(X_e\) (not the entire \(X\)) to avoid leakage. The generated \(P\) is a sequence of triplets of the form \((k_y^i, k_c^i, k_x^i)\) for \(i=1, \dots, n\), where \(\{k_y^i\}\) are parsed sub-SQL components from \(C_i\), \(\{k_c^i\}\) are the CoT reasoning content, and \(\{k_x^i\}\) are the corresponding natural language questions.

In the inner loop, we apply CoT to each parsed SQL component. The \textsc{Parser} focuses on tuple semantic shifts by tracking how each operation changes tuple semantics, which helps to interpret SQL more accurately. For each parsed sub-SQL \(k_y^i\), the corresponding question is generated by:

\vspace{-0.2cm}
\begin{equation}\small
(k_x^t, P^{>t}) = \textsc{Parser}(C_i, S, X_e, P^{< t}, k_y^t, k_c^t),
\end{equation}
where \(P^{< t}\) is the previously generated \((k_x, k_c, k_y)\) for \(C_i\) up to step \(t\).

\vspace{-10pt}
\subsubsection{Tuple-Semantic Tracking}

In the inner loop of the nested CoT process, we use a tuple-semantic tracking method to generate correct knowledge units. For the molecular database in Figure~\ref{fig_example}, each tuple of \sql{MOLECULE} represents information about a molecule, while after \sql{ATOM INNER JOIN MOLECULE} operation, each tuple represents detailed information about an atom along with its associated molecule information. This shift in tuple semantics directly affects the meaning of aggregation operations: \sql{COUNT(ATOM.id)} counts atoms and is equivalent to \sql{COUNT(*)}, while \sql{COUNT(DISTINCT ATOM.molecule\_id)} counts molecules.

In \sys, this tuple-semantic tracking approach monitors how each SQL operation shifts tuple semantics. For each \(k_y^t\), the \textsc{Parser} infers the semantics of tuple step by step:

\vspace{-0.2cm}
\begin{equation}\small
(k_c^t, k_x^t, P^{>t}) = \textsc{Parser}(C_i, S, X_e, P^{<t}, k_y^t),
\end{equation}
where \(k_c^t\) is the CoT reasoning content for tracking tuple semantics, and \(k_x^t\) is the question generated from \(k_c^t\). This incremental inference process improves translation accuracy by ensuring consistency with the expected tuple semantics.


For example, the \(k_c^t\) of the first SQL in Figure~\ref{fig_example}:
\begin{lstlisting}[language=TXT]
This query counts the number of sodium atoms that are part of non-carcinogenic molecules. 
The INNER JOIN connects the atom and molecule tables based on their shared molecule_id, ensuring that only sodium atoms from non-carcinogenic molecules are included in the count.The WHERE clause filters for both sodium atoms and non-carcinogenic labels, and the COUNT function aggregates these results into a single value, reflecting the total number of sodium atoms in non-carcinogenic molecules.
\end{lstlisting}



\vspace{-10pt}
\subsubsection{Knowledge De-duplication}

We merge all parsed knowledge units from \(\mathcal{C}'\) into a unified set \(\mathcal{K}_0\). Because \(\mathcal{K}_0\) may contain duplicates or non-informative entries, we perform a de-duplication step. For each knowledge unit \(k_y^i\) in \(\mathcal{K}_0\), we check whether its execution result is identical to that of any earlier unit \(k_y^{<i}\) or is \sql{NULL}. If so, we remove \(k_y^i\) from \(\mathcal{K}_0\). The final de-duplicated set of knowledge units is \(\mathcal{K}\).


\vspace{-10pt}

\subsection{Knowledge Validation}

In the forward Text-to-SQL translation process, LLMs might produce incorrect SQL due to misunderstandings of \(S\). We adopt a backward SQL-to-Text approach (Section~\ref{sec_knowledge_parsing}) to annotate and interpret SQL semantics. However, the complexity of database semantics can still lead to incorrect knowledge units. While \cite{GAR} uses human annotation to avoid such errors, this approach adds cost. We propose a probability-based filter to improve the reliability of generated knowledge units.


\vspace{-10pt}
\subsubsection{Probability-based Filter}

We use a probability-based filter to improve the quality of knowledge units. For each \(k^t \in \mathcal{K}\) in \(\mathcal{M}\),the probability is calculated as:
\vspace{-0.2cm}
\begin{equation}\small
\begin{aligned}
p(k^t \mid S, X) &= p(k^t \mid C_i, S, X) \cdot p(C_i \mid S, X) \\
&= p(k^t \mid C_i, S, X_e) \cdot p(C_i \mid S, X),
\end{aligned}
\end{equation}
where we assume each knowledge unit \(k^t\) comes from a unique SQL \(C_i\), and each question \(X\) has unique evidence \(X_e\).


We approximate by ignoring \(p(k^t \mid C_i, S, X_e)\) because the nested CoT process reduces the chance of generating duplicate knowledge units, making \(p(k^t \mid C_i, S, X_e)\) effectively constant. We then compute \(p(C_i \mid S, X)\) from the probability of generating the same execution result, rather than matching the output token sequence, to avoid sparsity in exact SQL matching.


We remove knowledge units whose probability is below a threshold \(\tau_0\):
\vspace{-0.1cm}
\begin{equation}\small
p(C_i \mid S, X) < \tau_0.
\vspace{-0.1cm}
\end{equation}
This step prevents low-quality knowledge units from entering \(\mathcal{M}\).


\vspace{-10pt}
\subsection{Knowledge-Enhanced Translation}

In the third step, we generate SQL based on \(\mathcal{M}\). This process can be viewed as an in-domain ICL Text-to-SQL translation because the demonstrations come from the same database. Using domain-specific demonstrations boosts performance, and schema information is provided only once since all examples reference the same database. For the final SQL generation, we apply multi-path generation with a self-consistency strategy to ensure both robustness and accuracy.


\vspace{-10pt}
\subsubsection{In-domain Demonstration Selection}

In this step, we aim to identify the relevant knowledge units from the memory to serve as demonstrations. To ensure the retrieval of database-specific knowledge that may not be effectively captured by structural or syntactic similarity measures, we focus exclusively on semantic similarity. Specifically, we select the demonstrations exhibiting the highest semantic alignment and compute the similarity between the question \(X\) and a knowledge unit \(k^i\) as:



\vspace{-0.2cm}
\begin{equation}\small
\text{sim}(X, k^i) = \cos\big(\text{EMB}(X), \text{EMB}(k^i_x)\big),
\end{equation}
where \(\text{EMB}\) converts sentences into vector representations. We implement \(\text{EMB}\) with Sentence-BERT and use FAISS to enable fast demonstration selection. A knowledge unit example is shown as:
\begin{lstlisting}[language=JSON]
Question: How many sodium atoms are found in non-carcinogenic molecules?
SQL: SELECT COUNT(T1.atom_id) FROM atom AS T1 INNER JOIN molecule AS T2 ON T1.molecule_id = T2.molecule_id WHERE T1.element = 'na' AND T2.label = '-';
Exec_result: [[17]]
\end{lstlisting}
\textit{Exec\_result} is the SQL execution result. For large results, we retain only the top 3 tuples in the knowledge unit.

\vspace{-10pt}
\subsubsection{Text-to-SQL Generation}

In the final Text-to-SQL step, we do not include the entire schema in the prompt. This schema linking strategy reduces the prompt length and inference cost for the \textsc{Coder} while focusing on relevant schema items eases the inference process. For each schema item \(s_i \in \mathcal{S}\), we include \(s_i\) only if:
\begin{equation}\small
s_i \in \bigcup_{k \in \mathcal{D}} \{ \text{ES}(k_y) \},
\vspace{-0.2cm}
\end{equation}
where \(\text{ES}\) extracts all schema items used in all of the input SQL.


Although \sys may rely on selected demonstrations \(\mathcal{D}\) for schema linking and SQL translation, it performs well due to the coverage and reliability of \(\mathcal{M}\). The knowledge parsing process includes all system logs, which contain candidate SQL for the current question, ensuring sufficient coverage. The \textsc{Validator} maintains the trustworthiness of \(\mathcal{M}\). We formulate the prompt according to the template in Figure~\ref{fig_overview} and use multi-path generation with a self-consistency mechanism to select the most reliable SQL output.


\section{Experiments}



\subsection{Experimental Setup}

\textbf{Datasets.}~~We evaluate \sys on three datasets: \textsc{Bird}~\cite{BIRD}, \textsc{Spider}~\cite{SPIDER}, and \textsc{Science}~\cite{SCIENCE}.


\textsc{\textsc{Bird}} includes 12{,}751 Text-to-SQL pairs from 95 large-scale databases across 37 professional fields, addressing noisy database values and leverages external knowledge for SQL generation.


\textsc{\textsc{Spider}} contains 10{,}181 natural language questions and 5{,}693 unique SQL from 206 databases spanning 138 domains for evaluation.


\textsc{\textsc{Science}} comprises three real-world scientific databases. Domain experts created 100/99/100 high-quality question-SQL pairs for each database.


\textbf{Evaluation.}~~We use execution accuracy (EX) as the primary evaluation metric. EX evaluates the accuracy of the SQL output by comparing the results of the predicted query with the gold query when executed on specific databases.

\textbf{Baselines.}~~We select several advanced prompting-based methods and compare \sys with these baseline models, including MAC-SQL~\cite{MAC}, DIN-SQL~\cite{DIN}, DAIL-SQL~\cite{DAIL}, PURPLE~\cite{purple}, CHESS~\cite{CHESS}, E-SQL~\cite{e-sql}, and RSL-SQL~\cite{rsl-sql}. Details of these balines are shown in Appendix~\ref{sec:baseline}. 



\textbf{Implementation Details.}~~Considering the trade-off between model performance and cost efficiency, we implement \sys and the baseline methods using GPT-4o-mini. The translation history utilized by \sys is generated by CHESS based on GPT-4o-mini. To further demonstrate the scalability of \sys for the base LLM, we conduct experiments with different foundation models, including Qwen2.5 Coder (Qwen2.5 Coder-14B/32B-Instruct), Qwen3 Coder (Qwen3-Coder-30B-A3B-Instruct) and the non-thinking mode of DeepSeek-V3 (DeepSeek-V3.2-Exp). We set the probability-based filter threshold to \(\tau = 0.3\) and use 30 demonstrations during the ICL SQL generation process.

\vspace{-10pt}
\subsection{Main Results}
\label{sec:main_res}

\begin{table*}[t]
\vspace{-5pt}
\centering
\caption{EX score (\%) on BIRD, SPIDER and SCIENCE dev datasets.}
\resizebox{\textwidth}{!}{
\begin{tabular}{|l||c|c|c|c||c|c|c|c|c||c|c|c|c|}
\hline
\multicolumn{1}{|c||}{\textbf{Method}} & \multicolumn{4}{c||}{\textbf{BIRD}} & \multicolumn{5}{c||}{\textbf{SPIDER}} & \multicolumn{4}{c|}{\textbf{SCIENCE}} \\
\cline{2-14}
& \scriptsize Sim. & \scriptsize Mod. & \scriptsize Chall. & \scriptsize Total & \scriptsize Easy & \scriptsize Med. & \scriptsize Hard & \scriptsize Ex.Hard & \scriptsize Total & \scriptsize CORDIS & \scriptsize ONCOMX & \scriptsize SDSS & \scriptsize Total \\
\hline
\hline
DIN-SQL & 52.43 & 31.61 & 25.69 & 43.61 & 83.9 & 79.1 & 68.4 & 60.2 & 75.4 & 51.00 & 52.53 & 7.00 & 36.79 \\
MAC-SQL & 60.11 & 46.67 & 35.42 & 53.72 & 91.1 & 83.4 & 66.1 & 78.4 & 78.4 & 51.00 & 55.56 & 14.00 & 40.13 \\
DAIL-SQL & 54.38 & 33.12 & 29.86 & 45.63 & 86.7 & 80.3 & 66.1 & 50.6 & 74.7 & 52.00 & 50.51 & 10.00 & 37.46 \\
E-SQL & 64.43 & 49.89 & 41.67 & 57.89 & 87.1 & 82.5 & 60.3 & 58.4 & 76.0 & - & - & - & - \\
RSI-SQL & 67.38 & 50.95 & 43.89 & 60.20 & 93.5 & 85.0 & 73.6 & 62.7 & 81.5 & 58.00 & 63.64 & 8.00 & 43.14 \\
PURPLE & 62.70 & 48.82 & 38.19 & 56.19 & \textbf{96.8} & \textbf{89.7} & \textbf{75.9} & \textbf{67.5} & \textbf{85.5} & 54.00 & 51.52 & \textbf{33.00} & 46.15 \\
CHESS$_{UT}$ & 69.08 & 52.69 & 45.15 & 61.86 & 91.2 & 83.5 & 69.9 & 55.2 & 78.5 & \textbf{62.00} & 65.66 & 18.00 & 48.89 \\
CHESS$_{V}$ & 69.92 & 52.69 & 43.06 & 61.99 & 91.5 & 83.6 & 69.5 & 56.0 & 78.7 & 55.00 & 58.59 & 14.00 & 42.47 \\
\hline
ORANGE & \textbf{71.24} & \textbf{57.20} & \textbf{51.39} & \textbf{65.12} & 91.1 & 87.7 & 73.6 & 60.8 & 81.8 & \textbf{62.00} & \textbf{70.71} & 31.00 & \textbf{54.52} \\
\hline
\end{tabular}
}
\label{tab_combined_results}
\vspace*{-10pt}
\end{table*}


We evaluated all methods using GPT-4o-mini, and results are shown in Table~\ref{tab_combined_results}. \sys achieves the best performance with an EX score of 65.12\%, outperforming the strongest baseline by 3.13\%. Notably, on the challenging Text-to-SQL tasks, \sys surpasses the best baseline by 6.24\%, demonstrating its effectiveness in handling complex queries. The improvement can be attributed to the domain-specific knowledge bases constructed by \sys. In-domain demonstrations enable more precise semantic alignment with the target database, while stored partial SQL semantics facilitate efficient completion of complex queries through knowledge reuse.

The candidate SQL generation step of \sys is based on CHESS, thus the comparison with CHESS highlights the advantage of \sys. \sys significantly outperforms both CHESS variants (by 3.13-3.26\%), showing that the knowledge parsing and reusing strategy provides greater benefits than SQL testing or voting strategies. Compared to other methods with the same base LLM, including CoT-based (DIN-SQL), multi-step reasoning (MAC-SQL), and out-domain methods (PURPLE), \sys also shows superior reasoning ability. Using the in-domain ICL strategy, \sys outperforms PURPLE and DAIL-SQL, which rely on SQL structural and semantic similarity correspondingly, demonstrating the in-domain demonstrations supply more relevant knowledge to assist the LLM in Text-to-SQL translation.

On the relatively simple \textsc{Spider} benchmark, \sys substantially improves over both CHESS variants, confirming its general reliability. While \sys doesn't achieve the best performance, this can be attributed to the reliance on CHESS for candidate generation and is further discussed in Section~\ref{sec:dependency}.

To evaluate the performance of \sys on more complex domain-specific scenarios, we employ \sys on the \textsc{Science} benchmark, which features three specialized domains and demands comprehensive semantic knowledge. \sys outperforms CHESS$_{UT}$ by 5.63\% and is the only method that exceeds an EX score of 50, highlighting its strong domain adaptability for specialized databases. As E-SQL involves numerous full data retrieval operations, which are infeasible on large-scale databases in \textsc{Science}, its experimental result is not reported.


\vspace{-10pt}
\subsection{Hyper-parameter Analysis}

\begin{figure}[t]
\vspace{-15pt} 
    \centering
    \includegraphics[width=0.63\linewidth]{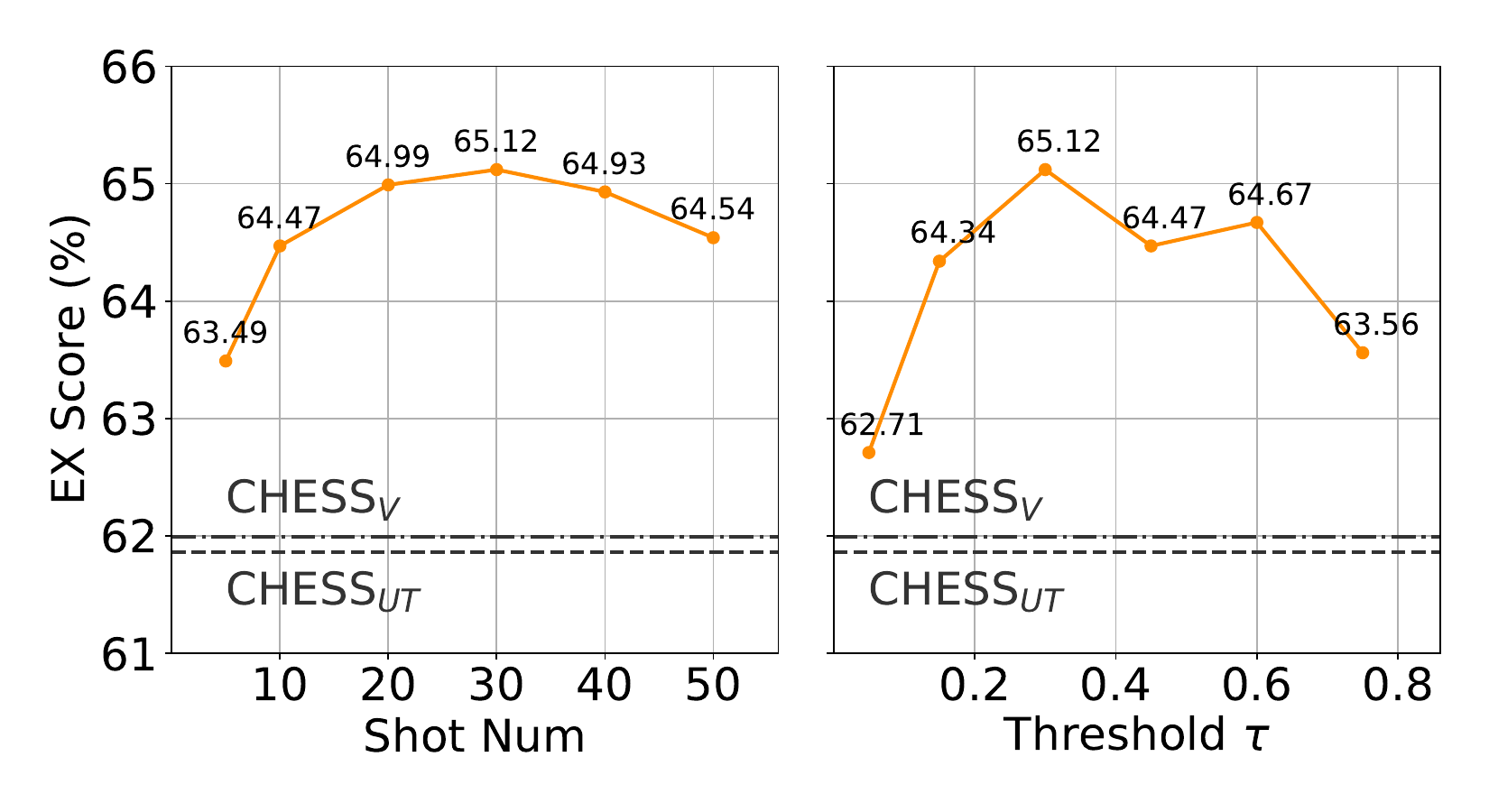}   
    \vspace{-10pt} 
    \caption{EX score (\%) on \textsc{Bird} dev under various hyper-parameters of \sys.}
    \vspace{-15pt} 
\label{fig_hyper}
\end{figure}

We analyze two key hyper-parameters in \sys: the number of in-context demonstrations (\textit{shot num}) and the probability threshold \(\tau\) for knowledge filtering in the probability-based filter. We set $\textit{shot num}=30$ and $\tau=0.3$ by default and vary one parameter while fixing the other. Figure~\ref{fig_hyper} illustrates how performance varies under different configurations.

The left plot shows the EX score trend under different \textit{shot num} values. The EX score of \sys initially increases as \textit{shot num} grows but eventually decreases. This pattern reflects a trade-off: too few shots may omit demonstrations with useful knowledge, while too many introduce irrelevant examples, adding noise and leading to a performance drop.

Similarly, the right plot indicates that the EX score first rises with stricter filtering (higher $\tau$) but eventually decreases. A low threshold retains incorrect knowledge units, whereas an overly high one discards semantically correct units, both degrading translation quality.

\vspace{-5pt}
\subsection{Ablation Study}



To evaluate the contributions of different modules in \sys, we conduct the ablation study. 

\begin{wraptable}[7]{r}{5.4cm}
\vspace*{-3\baselineskip}%
\scriptsize
\centering 
\caption{Ablation study of ORANGE.}
\label{tab:ablation}
\vspace{2mm}
\begin{tabular}{lc}
\toprule
\textbf{Strategy} & \textbf{EX Score (\%)} \\
\midrule
\sys & 65.12 \\
- History & 63.62 (-1.50) \\
- Validator & 62.71 (-2.41) \\
- Ranking & 62.32 (-2.80) \\
- ALL & 61.99 (-3.13) \\
- Schema Linking & 64.47 (-0.65) \\
\bottomrule
\end{tabular}
\end{wraptable}

\textit{--~History} removes knowledge from other historical translation tasks in the same database and only use candidates for the current task to generate knowledge units. The performance degradation observed in the absence of knowledge from other tasks highlights the self-evolution capability of \sys. 

\textit{--~Validator} removes the validator but retains all knowledge units. The notable performance decrease demonstrates the importance of knowledge unit validation, which enables \sys to exclude incorrect knowledge units.

\textit{--~Ranking} replaces the ranking-based demonstration selection with random sampling, resulting in a 2.8\% performance loss, indicating that selecting relevant knowledge units as demonstrations is crucial for enhancing in-domain translation accuracy of \sys.

\textit{--~ALL} removes the entire \sys framework and uses only the voting results (equal to CHESS$_V$). This ablation shows a performance drop of 3.13\%, emphasizing the comprehensive impact of the \sys strategy.

\textit{--~Schema Linking} causes a minor performance decline. Nonetheless, the schema linking strategy is still an appropriate design, as it improves performance while reducing the cost for LLMs.


\vspace{-0.2cm}
\subsection{Dependency on Prior Generation Process}
\label{sec:dependency}



As discussed in Section~\ref{sec:main_res}, while \sys achieves strong performance on complex, domain-specific benchmarks such as \textsc{Bird} and \textsc{Science}, the results on the \textsc{Spider} dataset are constrained. This limitation is primarily caused by the reliance on the quality of candidate SQL generation process. Due to the suboptimal performance of CHESS on the \textsc{Spider} dataset, the performance of \sys is accordingly constrained. 


\begin{table}[t]
\vspace{-5pt}
\caption{EX score (\%) of various methods with different prior SQL generator on \textsc{Bird}, \textsc{Spider} and \textsc{Science} dev datasets.}
\centering
\begin{tabular}{|l@{\hspace{7pt}}||c@{\hspace{5pt}}c||c@{\hspace{7pt}}c||c@{\hspace{3pt}}c|}
\hline
\multirow{2}{*}{\textbf{Method}} & \multicolumn{2}{c||}{\textbf{BIRD}} & \multicolumn{2}{c||}{\textbf{SPIDER}} & \multicolumn{2}{c|}{\textbf{SCIENCE}} \\
\cline{2-7}
 & \scriptsize EX & \scriptsize Diff. & \scriptsize EX & \scriptsize Diff. & \scriptsize EX & \scriptsize Diff. \\
\hline
\hline
PURPLE & 56.19 & {\scriptsize \textcolor{red!80!black}{(-8.09)}} & 85.5 & {\scriptsize \textcolor{red!80!black}{(-0.3)}} & 46.15 & {\scriptsize \textcolor{red!80!black}{(-7.03)}} \\
PURPLE+ORANGE & 64.28 & & 85.8 & & 53.18 & \\
\hline
CHESS$_{UT}$ & 61.86 & {\scriptsize \textcolor{red!80!black}{(-3.26)}} & 78.5 & {\scriptsize \textcolor{red!80!black}{(-3.3)}} & 48.49 & {\scriptsize \textcolor{red!80!black}{(-6.03)}} \\
CHESS$_{V}$ & 61.99 & {\scriptsize \textcolor{red!80!black}{(-3.13)}} & 78.7 & {\scriptsize \textcolor{red!80!black}{(-3.1)}} & 42.47 & {\scriptsize \textcolor{red!80!black}{(-12.05)}} \\
CHESS+ORANGE & 65.12 & & 81.8 & & 54.52 & \\
\hline
\end{tabular}
\label{dependency_result}
\vspace{-10pt}
\end{table}

To further evaluate the effectiveness of \sys and assess its robustness across different prior generation models, we conduct an additional experiment using PURPLE as the candidate SQL generator. As shown in Table~\ref{dependency_result}, when integrated with PURPLE, \sys achieves a new state-of-the-art on the \textsc{Spider} dataset, improving from 85.5\% to 85.8\%. More impressively, on the \textsc{Bird} and \textsc{Science} datasets, the PURPLE+\sys combination leads to even more substantial performance gains, demonstrating the robustness of \sys under different prior generation conditions.



\vspace{-10pt}
\subsection{Self-Evolutionary Performance}

To explore the long-term learning capacity and model scalability of \sys, we examine its performance under various historical knowledge settings. We simulate three deployment scenarios: (1) Self-Only Context, (2) Accumulated History, and (3) All History and report the number of available knowledge units under each scenario.


\textit{Self-Only Context} acts as a cold-start situation without prior knowledge, corresponding to the baseline performance of \sys without knowledge accumulation. Only SQL candidates from the current translation task are available for ICL demonstration selection.

\textit{Accumulated History} simulates real-world scenarios, where knowledge is incrementally obtained from sequentially processed tasks. Each translation can only utilize the knowledge derived from the completed tasks.

\textit{All History} is the default setting of \sys, using the full translation history of the target database to provide comprehensive knowledge coverage, which showcases the performance of \sys with ample historical data.






\begin{table}[t]
\caption{EX score (\%) on \textsc{Bird} dev and knowledge units statistics of \sys under different historical knowledge settings.}
\small
\centering
\begin{tabular}{|l||c@{\hspace{0.2cm}}c@{\hspace{0.2cm}}c@{\hspace{0.2cm}}c||c@{\hspace{0.2cm}}c@{\hspace{0.2cm}}c|}
\hline
\multirow{2}{*}{\textbf{History}} & \multicolumn{4}{c||}{\textbf{EX score (\%) on BIRD dev}} & \multicolumn{3}{c|}{\textbf{KU Count}} \\
\cline{2-8}
 & \scriptsize Simple & \scriptsize Moderate & \scriptsize Challenging & \scriptsize Total & \scriptsize Average & \scriptsize Min & \scriptsize Max \\
\hline
\hline
Self-Only Context & 70.49 & 54.41 & 49.31 & 63.62 & 3.01 & 0 & 11 \\
Accumulated History & 70.49 & 56.77 & 50.69 & 64.47 & 228.8 & 1 & 617 \\
All History & 71.24 & 57.20 & 51.39 & 65.12 & 459.3 & 222 & 617 \\
\hline
\end{tabular}
\label{tab_combined_history_knowledge}
\end{table}

As shown in Table~\ref{tab_combined_history_knowledge}, \sys demonstrates clear evolutionary improvement as historical knowledge accumulates. \textit{All History} achieves the best performance, and even \textit{Self-Only Context} outperforms other baseline methods. This progressive enhancement highlights the the long-term adaptability and self-evolution capability of \sys through continuous knowledge integration, offering a scalable advantage in real-world applications.

\vspace{-10pt}
\subsection{Scalability with Different Foundation Models}


To assess the architectural independence and scalability of our framework, we conducted experiments with different foundation models.


\begin{table}[t]
\caption{EX score(\%) on \textsc{Bird} dev with different base models of \sys.}
\label{tab:base_model_compare}
\centering
\small
\begin{tabular}{|l@{\hspace{0.2cm}}||c@{\hspace{0.15cm}}c@{\hspace{0.15cm}}c@{\hspace{0.15cm}}c|@{\hspace{0.1cm}}}
\hline
\textbf{Base Model} & \textbf{Simple} & \textbf{Moderate} & \textbf{Challenging} & \textbf{Total} \\
\hline
\hline
GPT-4o-mini & 71.24 & 57.20 & 51.39 & 65.12 \\
\hline
Qwen2.5-Coder-14B-Instruct & 74.16 & 56.90 & 54.48 & 67.08 \\
\hline
Qwen2.5-Coder-32B-Instruct & 73.19 & 60.99 & 52.41 & 67.54 \\
\hline
Qwen3-Coder-30B-A3B-Instruct & 74.38 & 60.78 & 55.86 & 68.51 \\
\hline
DeepSeek-V3.2-Exp & \textbf{76.11} & \textbf{61.21} & \textbf{62.07} & \textbf{70.27} \\
\hline
\end{tabular}
\vspace{-10pt}
\end{table}

As shown in Table~\ref{tab:base_model_compare}, \sys demonstrates consistent performance improvements across foundation models. While GPT-4o-mini registers at 65.12\% on the \textsc{Bird} dataset, Qwen family models reach up to 68.51\%, with DeepSeek-V3 further advancing to 70.27\%. The most notable gain occurs in challenging tasks, where DeepSeek-V3 shows a 10.68\% improvement over GPT-4o-mini, confirming its effectiveness in employing advanced models for complex reasoning.

\vspace{-5pt}
\section{Related Works}

Enhancing the reasoning process of LLMs is crucial for generating accurate SQL, especially for complex queries in the Text-to-SQL translation tasks.~\cite{pourreza2024din,liu2023divide,tai2023exploring} This is often achieved by decomposing complex questions into simpler, intermediate steps, as exemplified by Chain-of-Thought (CoT) prompting~\cite{wei2022chain,kojima2022large} and its variants like Least-to-Most Prompting~\cite{zhou2022least}. To enhance robustness, Self-consistency explores multiple reasoning paths and selects the most frequent answer through majority voting~\cite{wang2022self}. A related stream focuses on iterative refinement, where the model improves its output through self-correction. This can be guided by model-generated critique (Self-improvement)~\cite{touvron2023llama,zelikman2022star}, execution feedback from the database (Self-debugging)~\cite{andrew2024evaluating,CHESS}, or other generated auxiliary information~\cite{zheng2023progressive,welleck2022generating}. While these strategies improve the SQL translation performance, they can not acquire and accumulate past translation experiences and fail to equip the model with domain-specific insights.



Recognizing the limitations of relying solely on intrinsic reasoning, several researches highlight the utilization of in-domain knowledge. SQL-aligned demonstrations~\cite{gao2023text,poesia2022synchromesh,nan2023enhancing,purple} semantically similar examples \cite{an2023skill} are incorporated in In-context prompting. Some methods explicitly provide domain-specific instructions or demonstrations~\cite{dong2023c3,gu2023interleaving,liu2023comprehensive,arora2023adapt}. To overcome the limited number of demonstrations, other approaches leverage larger-scale knowledge sources, such as generating synthetic domain-specific question-SQL pairs~\cite{chang2023selective} or utilizing historical query logs~\cite{CHASE,LPE-SQL}. However, In-context demonstrations are constrained by the finite context windows of LLMs and struggle with knowledge scalability. Synthetic data generation inevitably fails to fully capture the real intent of user queries, and the generation process itself often introduces additional noise. Meanwhile, log-based approaches typically require extensive manual annotation or human interaction, thereby necessitating a more scalable, reliable, and automated mechanism for in-domain knowledge integration.

\vspace{-5pt}
\section{Conclusion}




We propose \sys, a self-evolutionary Text-to-SQL method that enhances complex reasoning by parsing and validating in-domain knowledge from translation history. Through continual memory updates after each translation, \sys expands its knowledge of database-specific semantics, progressively improving translation accuracy without human intervention. This approach offers a practical and scalable solution for deploying Text-to-SQL in real-world scenarios.

Several future directions are also envisioned. Since \sys is compatible with most Text-to-SQL methods, researchers can experiment with alternative SQL generation techniques during the cold-start phase. While our nested CoT strategy demonstrates notable accuracy improvement, it remains adaptable to various reasoning paradigms. Further advancements in LLM-based reasoning could refine the quality of knowledge generation and filtering, leading to more robust Text-to-SQL translation.


\appendix
\vspace{-5pt}

\section{Case Study}

We compare different models on the \textsc{Bird} dataset and analyze their outputs.

\begin{figure*}[t]
    \vspace{-5pt}
    \includegraphics[width=0.99\linewidth]{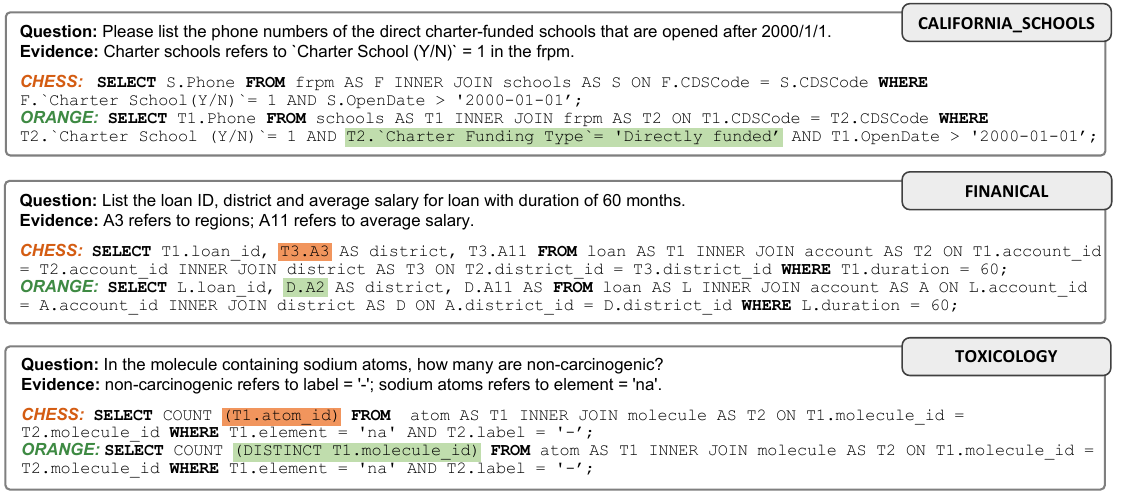}
    \vspace{-0.2cm}
    \caption{Case Study.}
    \vspace{-10pt}
    \label{fig_case_study}
\end{figure*}

Figure~\ref{fig_case_study} shows SQL queries generated by CHESS and \sys, along with their corresponding NL questions, for three databases: \sql{CALIFORNIA\_SCHOOLS}, \sql{FINANCIAL}, and \sql{TOXICOLOGY}. In these examples, \sys comprehends natural language semantics more accurately and generates more precise SQL.


In the \sql{CALIFORNIA\_SCHOOLS} database, CHESS fails to capture a critical semantic detail: the term \textit{direct} indicates that the funding type of schools should be \textit{Directly funded}. Consequently, when selecting schools from \sql{FRPM}, the query should include \sql{FRPM.`Charter Funding Type`='Directly funded'}, instead of using \sql{FRPM.`Charter School (Y/N)`=1} alone.

CHESS aligns questions with schema descriptions based solely on restrained individual evidence. As illustrated in the case of \sql{FINANCIAL}, due to the absence of explicit mention of \sql{A2}, CHESS incorrectly selects \sql{A3} (represents regions) for \textit{district}. In contrast, \sys can leverage its domain-specific knowledge and identify that \sql{A2} corresponds to districts and \sql{A3} to regions, thus correctly using \sql{A2} in the generated SQL.

Operations such as \sql{JOIN} and \sql{GROUP BY} can alter tuple semantics during aggregation. For instance, \sql{ATOM INNER JOIN MOLECULE} transforms each tuple to represent an atom associated with a specific molecule, rather than the atom alone. When counting atoms, \sys recognizes this semantic shift and uses \sql{molecule\_id}, while CHESS incorrectly uses \sql{atom\_id}.


\vspace{-5pt}
\section{Baseline Details}

\label{sec:baseline}

\textsc{\textbf{MAC-SQL}}~\cite{MAC} proposes a multi-step reasoning approach to address complex questions, which are decomposed into smaller, more manageable sub-questions and subsequently solved by different agents. To further enhance performance, the refiner agent employs external tools for SQL execution and iteratively refines faulty SQL queries according to the feedback.
 
\textsc{\textbf{DIN-SQL}}~\cite{DIN} incorporates task classification and problem decomposition to handle the complex SQL generation task.  It categorizes input questions into three distinct types based on the presence of sub-queries and multi-table \sql{JOIN} operations. For each category, tailored prompts are employed to reduce mismatch issues during SQL translation. DIN-SQL uses a standard in-context learning (ICL) framework without relying on explicit similarity-based retrieval metrics. 

\textsc{\textbf{DAIL-SQL}}~\cite{DAIL} focuses on the example selection and organization in few-shot prompting strategies. It employs DAIL Selection, a retrieval method that extracts demonstrations based on semantic similarity, considering both questions and queries to better align the retrieved demonstrations with the target query.

\textsc{\textbf{PURPLE}}~\cite{purple} tackles the difficulty of generating SQL queries involving complex logical operator compositions. To enhance the SQL-writing capabilities of LLMs, PURPLE masks specific values and highlights logical operations within SQL queries during demonstration selection. It adopts a retrieval strategy grounded in SQL structural similarity, enabling the model to better generalize to intricate SQL logic patterns.

\textsc{\textbf{CHESS}}~\cite{CHESS} adopts a pipeline that involves retrieving relevant entities and context, optimizing schema, generating SQL candidates, and ultimately selecting the final SQL from them. CHESS provides two SQL selection strategies: Unit Testing (denoted as CHESS$_{UT}$), which selects the query with the most consistent execution results, and Voting (denoted as CHESS$_{V}$), which involves multiple LLMs voting and ranking the candidates.

\textsc{\textbf{E-SQL}}~\cite{e-sql} integrates schema information directly into the question representation, rather than conducting dependent schema linking. This approach is claimed to effectively narrow the gap between natural language queries and database structures.

\textsc{\textbf{RSL-SQL}}~\cite{rsl-sql} seeks to balance the risks of overlooking important information in a complex schema and the inefficiencies of using a simplified schema. It generates SQL queries in two scenarios: one with the full schema and one with a simplified schema enriched by extra context and selects the final SQL from the generated candidates.

\bibliographystyle{splncs04}
\bibliography{main}

\begin{thebibliography}{10}
\providecommand{\url}[1]{\texttt{#1}}
\providecommand{\urlprefix}{URL }
\providecommand{\doi}[1]{https://doi.org/#1}

\bibitem{an2023skill}
An, S., Zhou, B., Lin, Z., Fu, Q., Chen, B., Zheng, N., Chen, W., Lou, J.G.: Skill-based few-shot selection for in-context learning. arXiv preprint arXiv:2305.14210  (2023)

\bibitem{andrew2024evaluating}
Andrew, J.J., Vincent, M., Burgun, A., Garcelon, N.: Evaluating llms for temporal entity extraction from pediatric clinical text in rare diseases context. In: Proceedings of the First Workshop on Patient-Oriented Language Processing (CL4Health)@ LREC-COLING 2024. pp. 145--152 (2024)

\bibitem{arora2023adapt}
Arora, A., Bhaisaheb, S., Nigam, H., Patwardhan, M., Vig, L., Shroff, G.: Adapt and decompose: Efficient generalization of text-to-sql via domain adapted least-to-most prompting. arXiv preprint arXiv:2308.02582  (2023)

\bibitem{e-sql}
Cafero{\u{g}}lu, H.A., Ulusoy, {\"O}.: E-sql: Direct schema linking via question enrichment in text-to-sql. arXiv preprint arXiv:2409.16751  (2024)

\bibitem{rsl-sql}
Cao, Z., Zheng, Y., Fan, Z., Zhang, X., Chen, W., Bai, X.: Rsl-sql: Robust schema linking in text-to-sql generation. arXiv preprint arXiv:2411.00073  (2024)

\bibitem{SQLFixAgent}
Cen, J., Liu, J., Li, Z., Wang, J.: Sqlfixagent: Towards semantic-accurate {SQL} generation via multi-agent collaboration. CoRR  \textbf{abs/2406.13408} (2024)

\bibitem{chang2023selective}
Chang, S., Fosler-Lussier, E.: Selective demonstrations for cross-domain text-to-sql. arXiv preprint arXiv:2310.06302  (2023)

\bibitem{chen2025reliable}
Chen, K., Chen, Y., Koudas, N., Yu, X.: Reliable text-to-sql with adaptive abstention. Proceedings of the ACM on Management of Data  \textbf{3}(1),  1--30 (2025)

\bibitem{SelfDebug}
Chen, X., Lin, M., Sch{\"{a}}rli, N., Zhou, D.: Teaching large language models to self-debug. In: The Twelfth International Conference on Learning Representations, {ICLR} 2024, Vienna, Austria, May 7-11, 2024. OpenReview.net (2024)

\bibitem{LPE-SQL}
Chu, Z., Wang, Z., Qin, Q.: Leveraging prior experience: An expandable auxiliary knowledge base for text-to-sql. arXiv preprint arXiv:2411.13244  (2024)

\bibitem{dong2023c3}
Dong, X., Zhang, C., Ge, Y., Mao, Y., Gao, Y., Lin, J., Lou, D., et~al.: C3: Zero-shot text-to-sql with chatgpt. arXiv preprint arXiv:2307.07306  (2023)

\bibitem{GAR}
Fan, Y., He, Z., Ren, T., Guo, D., Chen, L., Zhu, R., Chen, G., Jing, Y., Zhang, K., Wang, X.S.: Gar: {A} generate-and-rank approach for natural language to {SQL} translation. In: 39th {IEEE} International Conference on Data Engineering, {ICDE} 2023, Anaheim, CA, USA, April 3-7, 2023. pp. 110--122. {IEEE} (2023)

\bibitem{DAIL}
Gao, D., Wang, H., Li, Y., Sun, X., Qian, Y., Ding, B., Zhou, J.: Text-to-sql empowered by large language models: A benchmark evaluation. arXiv preprint arXiv:2308.15363  (2023)

\bibitem{gao2023text}
Gao, D., Wang, H., Li, Y., Sun, X., Qian, Y., Ding, B., Zhou, J.: Text-to-sql empowered by large language models: A benchmark evaluation. arXiv preprint arXiv:2308.15363  (2023)

\bibitem{xiyan}
Gao, Y., Liu, Y., Li, X., Shi, X., Zhu, Y., Wang, Y., Li, S., Li, W., Hong, Y., Luo, Z., et~al.: Xiyan-sql: A multi-generator ensemble framework for text-to-sql. arXiv preprint arXiv:2411.08599  (2024)

\bibitem{gu2023interleaving}
Gu, Z., Fan, J., Tang, N., Zhang, S., Zhang, Y., Chen, Z., Cao, L., Li, G., Madden, S., Du, X.: Interleaving pre-trained language models and large language models for zero-shot nl2sql generation. arXiv preprint arXiv:2306.08891  (2023)

\bibitem{knowledge-to-sql}
Hong, Z., Yuan, Z., Chen, H., Zhang, Q., Huang, F., Huang, X.: Knowledge-to-sql: Enhancing sql generation with data expert llm. arXiv preprint arXiv:2402.11517  (2024)

\bibitem{kojima2022large}
Kojima, T., Gu, S.S., Reid, M., Matsuo, Y., Iwasawa, Y.: Large language models are zero-shot reasoners. Advances in neural information processing systems  \textbf{35},  22199--22213 (2022)

\bibitem{BIRD}
Li, J., Hui, B., Qu, G., Yang, J., Li, B., Li, B., Wang, B., Qin, B., Geng, R., Huo, N., Zhou, X., Ma, C., Li, G., Chang, K.C., Huang, F., Cheng, R., Li, Y.: Can {LLM} already serve as {A} database interface? {A} big bench for large-scale database grounded text-to-sqls. In: Oh, A., Naumann, T., Globerson, A., Saenko, K., Hardt, M., Levine, S. (eds.) Advances in Neural Information Processing Systems 36: Annual Conference on Neural Information Processing Systems 2023, NeurIPS 2023, New Orleans, LA, USA, December 10 - 16, 2023 (2023)

\bibitem{PET-SQL}
Li, Z., Wang, X., Zhao, J., Yang, S., Du, G., Hu, X., Zhang, B., Ye, Y., Li, Z., Zhao, R., et~al.: Pet-sql: A prompt-enhanced two-round refinement of text-to-sql with cross-consistency. arXiv preprint arXiv:2403.09732  (2024)

\bibitem{liu2023comprehensive}
Liu, A., Hu, X., Wen, L., Yu, P.S.: A comprehensive evaluation of chatgpt's zero-shot text-to-sql capability. arXiv preprint arXiv:2303.13547  (2023)

\bibitem{liu2023divide}
Liu, X., Tan, Z.: Divide and prompt: Chain of thought prompting for text-to-sql. arXiv preprint arXiv:2304.11556  (2023)

\bibitem{nan2023enhancing}
Nan, L., Zhao, Y., Zou, W., Ri, N., Tae, J., Zhang, E., Cohan, A., Radev, D.: Enhancing few-shot text-to-sql capabilities of large language models: A study on prompt design strategies. arXiv preprint arXiv:2305.12586  (2023)

\bibitem{poesia2022synchromesh}
Poesia, G., Polozov, O., Le, V., Tiwari, A., Soares, G., Meek, C., Gulwani, S.: Synchromesh: Reliable code generation from pre-trained language models. arXiv preprint arXiv:2201.11227  (2022)

\bibitem{CHASE}
Pourreza, M., Li, H., Sun, R., Chung, Y., Talaei, S., Kakkar, G.T., Gan, Y., Saberi, A., Ozcan, F., Arik, S.{\"{O}}.: {CHASE-SQL:} multi-path reasoning and preference optimized candidate selection in text-to-sql. CoRR  \textbf{abs/2410.01943} (2024)

\bibitem{DIN}
Pourreza, M., Rafiei, D.: {DIN-SQL:} decomposed in-context learning of text-to-sql with self-correction. In: Oh, A., Naumann, T., Globerson, A., Saenko, K., Hardt, M., Levine, S. (eds.) Advances in Neural Information Processing Systems 36: Annual Conference on Neural Information Processing Systems 2023, NeurIPS 2023, New Orleans, LA, USA, December 10 - 16, 2023 (2023)

\bibitem{pourreza2024din}
Pourreza, M., Rafiei, D.: Din-sql: Decomposed in-context learning of text-to-sql with self-correction. Advances in Neural Information Processing Systems  \textbf{36} (2024)

\bibitem{purple}
Ren, T., Fan, Y., He, Z., Huang, R., Dai, J., Huang, C., Jing, Y., Zhang, K., Yang, Y., Wang, X.S.: Purple: Making a large language model a better sql writer. arXiv preprint arXiv:2403.20014  (2024)

\bibitem{tai2023exploring}
Tai, C.Y., Chen, Z., Zhang, T., Deng, X., Sun, H.: Exploring chain-of-thought style prompting for text-to-sql. arXiv preprint arXiv:2305.14215  (2023)

\bibitem{CHESS}
Talaei, S., Pourreza, M., Chang, Y., Mirhoseini, A., Saberi, A.: {CHESS:} contextual harnessing for efficient {SQL} synthesis. CoRR  \textbf{abs/2405.16755} (2024)

\bibitem{touvron2023llama}
Touvron, H., Martin, L., Stone, K., Albert, P., Almahairi, A., Babaei, Y., Bashlykov, N., Batra, S., Bhargava, P., Bhosale, S., et~al.: Llama 2: Open foundation and fine-tuned chat models. arXiv preprint arXiv:2307.09288  (2023)

\bibitem{MAC}
Wang, B., Ren, C., Yang, J., Liang, X., Bai, J., Zhang, Q., Yan, Z., Li, Z.: {MAC-SQL:} {A} multi-agent collaborative framework for text-to-sql. CoRR  \textbf{abs/2312.11242} (2023)

\bibitem{wang2024improving}
Wang, D., Dou, L., Zhang, X., Zhu, Q., Che, W.: Improving demonstration diversity by human-free fusing for text-to-sql. arXiv preprint arXiv:2402.10663  (2024)

\bibitem{wang2022self}
Wang, X., Wei, J., Schuurmans, D., Le, Q., Chi, E., Narang, S., Chowdhery, A., Zhou, D.: Self-consistency improves chain of thought reasoning in language models. arXiv preprint arXiv:2203.11171  (2022)

\bibitem{wei2022chain}
Wei, J., Wang, X., Schuurmans, D., Bosma, M., Xia, F., Chi, E., Le, Q.V., Zhou, D., et~al.: Chain-of-thought prompting elicits reasoning in large language models. Advances in neural information processing systems  \textbf{35},  24824--24837 (2022)

\bibitem{welleck2022generating}
Welleck, S., Lu, X., West, P., Brahman, F., Shen, T., Khashabi, D., Choi, Y.: Generating sequences by learning to self-correct. arXiv preprint arXiv:2211.00053  (2022)

\bibitem{DEA-SQL}
Xie, Y., Jin, X., Xie, T., Lin, M., Chen, L., Yu, C., Cheng, L., Zhuo, C., Hu, B., Li, Z.: Decomposition for enhancing attention: Improving llm-based text-to-sql through workflow paradigm. arXiv preprint arXiv:2402.10671  (2024)

\bibitem{SPIDER}
Yu, T., Zhang, R., Yang, K., Yasunaga, M., Wang, D., Li, Z., Ma, J., Li, I., Yao, Q., Roman, S., Zhang, Z., Radev, D.R.: Spider: {A} large-scale human-labeled dataset for complex and cross-domain semantic parsing and text-to-sql task. In: Riloff, E., Chiang, D., Hockenmaier, J., Tsujii, J. (eds.) Proceedings of the 2018 Conference on Empirical Methods in Natural Language Processing, Brussels, Belgium, October 31 - November 4, 2018. pp. 3911--3921. Association for Computational Linguistics (2018)

\bibitem{zelikman2022star}
Zelikman, E., Wu, Y., Mu, J., Goodman, N.: {ST}ar: Bootstrapping reasoning with reasoning. In: Oh, A.H., Agarwal, A., Belgrave, D., Cho, K. (eds.) Advances in Neural Information Processing Systems (2022)

\bibitem{zhang2024sqlfuse}
Zhang, T., Chen, C., Liao, C., Wang, J., Zhao, X., Yu, H., Wang, J., Li, J., Shi, W.: Sqlfuse: Enhancing text-to-sql performance through comprehensive llm synergy. arXiv preprint arXiv:2407.14568  (2024)

\bibitem{SCIENCE}
Zhang, Y., Deriu, J., Katsogiannis{-}Meimarakis, G., Kosten, C., Koutrika, G., Stockinger, K.: Sciencebenchmark: {A} complex real-world benchmark for evaluating natural language to {SQL} systems. Proc. {VLDB} Endow.  \textbf{17}(4),  685--698 (2023)

\bibitem{zheng2023progressive}
Zheng, C., Liu, Z., Xie, E., Li, Z., Li, Y.: Progressive-hint prompting improves reasoning in large language models. arXiv preprint arXiv:2304.09797  (2023)

\bibitem{zhou2022least}
Zhou, D., Sch{\"a}rli, N., Hou, L., Wei, J., Scales, N., Wang, X., Schuurmans, D., Cui, C., Bousquet, O., Le, Q., et~al.: Least-to-most prompting enables complex reasoning in large language models. arXiv preprint arXiv:2205.10625  (2022)

\end{thebibliography}

%






\end{document}